\documentclass[letter, traditabstract]{aa}

\usepackage{graphics}
\usepackage{xspace}
\usepackage{natbib}
\usepackage{longtable}
\usepackage{lscape}
\usepackage{txfonts}

\def\aap{A\&A}

\def\apjl{ApJL}

\def\um{$\mu$m\xspace}

\def\12c{$^{12}$C\xspace}
\def\h2{H$_2$\xspace}
\def\h2o{H$_2$O\xspace}
\def\so2{SO$_2$\xspace}
\def\co2{CO$_2$\xspace}
\def\co{CO\xspace}

\def\c2h2{C$_2$H$_2$\xspace}

\def\tio2{TiO$_2$\xspace}
\def\tio{TiO\xspace}

\begin{document}

   \title{Discovery of a TiO emission band in the infrared spectrum of
the S star NP Aurigae}
   \author{K. Smolders 
          \inst{1} \fnmsep \thanks{Aspirant Fellow of the Fund for Scientific Research, Flanders}
          \and
          T.~Verhoelst\inst{1,2}
          \and
          P.~Neyskens\inst{3} \fnmsep \thanks{Fellowship ``boursier F.R.I.A'', Belgium}
          \and
          J.A.D.L.~Blommaert\inst{1}
          \and
          L.~Decin\inst{1} 
          \and
          H.~Van~Winckel\inst{1}
          \and
          S.~Van~Eck\inst{3}
          \and
          G.~C.~Sloan\inst{4}
          \and
          J.~Cami\inst{5, 6}
          \and
          S.~Hony\inst{7}
          \and
          P.~De~Cat\inst{8}
          \and
          J.~Menu\inst{1}\fnmsep $^\star$
          \and
          J.~Vos\inst{1}
          }

   \offprints{K. Smolders}

   \institute{Instituut voor Sterrenkunde (IvS),
              Katholieke Universiteit Leuven,
              Celestijnenlaan 200 D,
              B-3001 Leuven, Belgium
         \and
              Belgian Institute for Space Aeronomy (BIRA-IASB),
              Ringlaan-3-Avenue Circulaire, B-1180 Brussels, Belgium
         \and
              Institut d'Astronomie et d'Astrophysique (IAA), Universit\'e 
              Libre de Bruxelles, C.P.226, Boulevard du Triomphe,
              B-1050 Bruxelles, Belgium
         \and
              Cornell University,
              Astronomy Department,
              Ithaca, NY 14853, USA
         \and
              Department of Physics and Astronomy,
              University of Western Ontario, 
              London, Ontario N6A 3K7, Canada
         \and
	          SETI Institute, 
	          189 Bernardo Ave, Suite 100, 
	          Mountain View, CA 94043, USA
 		\and
              Service d'Astrophysique,
              CEA Saclay,
              Bat.709 Orme des Merisiers,
              91191 Gif-sur-Yvette, France
         \and
			Koninklijke Sterrenwacht van Belgi\"e, 
			Ringlaan 3, 
			B-1180 Brussel, Belgium
             }
              
\date{\today}

   \abstract{We report on the discovery of an infrared emission band in the Spitzer spectrum of the S-type AGB star NP Aurigae that is caused by TiO molecules in the circumstellar environment. We modelled the observed emission to derive the temperature of the TiO molecules ($\approx 600$\,K), an upper limit on the column density ($\approx$ 10$^{17.25}$\,cm$^{-2}$) and a lower limit on the spatial extent of the layer that contains these molecules. ($\approx$\,4.6\,R$_{\star}$). This is the first time that this TiO emission band is observed. A search for similar emission features in the sample of S-type stars yielded two additional candidates. However, owing to the additional dust emission, the identification is less stringent. By comparing the stellar characteristics of NP~Aur to those of the other stars in our sample, we find that all stars with TiO emission show large-amplitude pulsations, s-process enrichment, and a low C/O ratio. These characteristics might be necessary requirements for a star to show TiO in emission, but they are not sufficient.}


\maketitle

\section{Introduction}
The spectral class of S stars contains objects that show absorption bands from oxides such as ZrO, LaO and YO in addition to the absorption bands of TiO that characterize M stars \citep{Merrill1922, Keenan1954}. It is often said that S-type stars have C/O ratios close to one. Although the ZrO bands become more pronounced when the C/O ratio is closer to unity, a star can show ZrO absorption bands while the C/O ratio is as low as 0.5 \citep{VanEck2010}. This is only possible if the stars have enhanced abundances of s-process elements such as Zr, La and Y. If this s-process enrichment is caused by nucleosynthesis and dredge-up on the thermally-pulsing AGB, these stars are \emph{intrinsic} S stars. \emph{Extrinsic} S stars are enriched through pollution by a binary companion \citep{Groenewegen1993, VanEck1999}.

The infrared spectra of S stars are diverse and show a mix of dust species typical for either oxygen-rich or carbon-rich circumstellar environments, molecular emission bands, or no excess emission at all \citep{Chen1993, Hony2009, Smolders2010, Smolders2012}. This heterogeneity of the infrared appearance within one spectral class is caused by the peculiar chemical composition of the stellar atmosphere of S stars. In oxygen-rich stars, the stable CO molecule forms at high temperatures in the stellar outflows until almost all free carbon is consumed, leaving only the free oxygen atoms to produce molecules and dust. For carbon-rich stars, the oxygen is depleted and only carbon is left to form molecules or dust grains. This CO dichotomy explains the differences in spectral appearance \citep{Millar2000, Willacy1997, Treffers1974}. Because S-type stars span a range of C/O ratios from near solar values to near unity, and since the composition of the molecular shell and dusty wind is highly sensitive to the actual C/O ratio, a rich diversity in gas and solid-state species can be expected \citep{Smolders2012}.

The results discussed in this paper are part of a program to study a large sample of S-type AGB stars (Program ID 30737, P.I. S.\,Hony), observed with the Infrared Spectrograph \citep[IRS, ][]{Houck2004} onboard the Spitzer Space Telescope \citep{Werner2004}. The sample was selected from the Stephenson S star catalog \citep{stephenson2}, limited to those targets that are observable with Spitzer based on the 2MASS and IRAS data. The extrinsic S stars were excluded from the sample based on the amplitude of variability \citep{Smolders2012}. The final sample consists of 87 S-type stars in total, containing 32 stars with dust emission features, 3 stars with only molecular emission features, 4 stars with hydrocarbon emission, and 47 stars without significant emission features. In addition to the infrared spectra, we acquired high-resolution optical spectroscopy using the HERMES \'{e}chelle spectrograph at the Mercator telescope in La Palma \citep{Raskin2011}, covering 3770-9000\,\AA\ with a resolving power of $\mathrm{R} \approx 85\,000$. \citet{Smolders2012} provide more details on the data and the sample selection.

Previous studies of this sample have already led to the identification of the fundamental ro-vibrational band of SiS at 13\,\um and its first overtone at 6.7\,\um in absorption \citep{Cami2009} and in emission \citep{Sloan2011}. Furthermore, four stars in this sample show the typical hydrocarbon emission features on top of an oxygen-rich photosphere \citep{Smolders2010}. In this letter we focus on NP\,Aurigae, which shows a double peaked emission feature in the 9.8--10.5~\um region, which we identify as the fundamental vibrational band of TiO. In Sects. 2 and 3 we present the identification and modeling of the emission feature. In Sect. 4, we discuss the search for TiO and TiO$_2$ in circumstellar environments and we show the tentative detection of TiO in two other S-type stars. Finally, we conclude in Sect. 5 with the implications of this finding.

\section{Identification of the TiO emission}
\begin{figure}
\resizebox{\hsize}{!}{\includegraphics{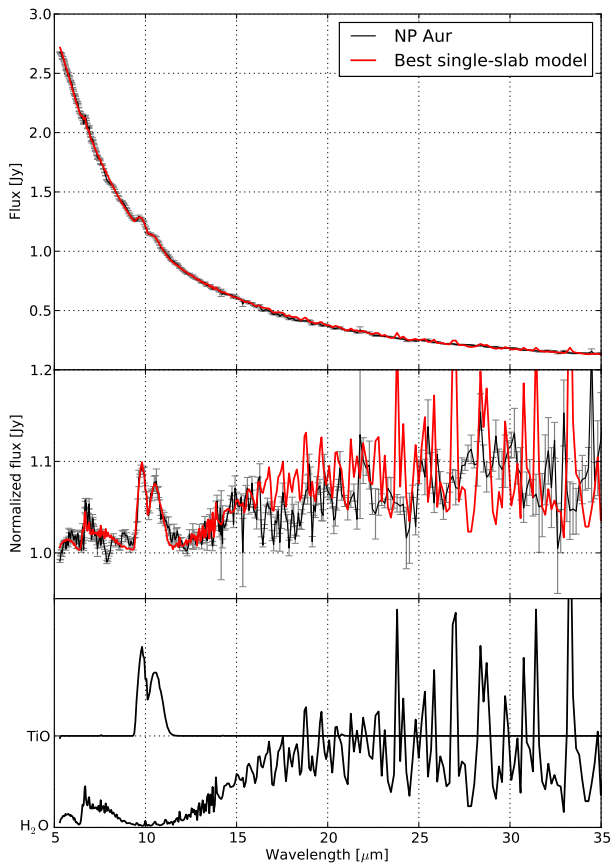}}
  \caption{Top panel: the infrared spectrum of NP~Aur shown in black and the slab model, shown in red; middle panel: spectrum and model, divided by the blackbody radiation of the stellar disk; bottom panel: the contribution of TiO and H$_2$O separately.}
  \label{fig:fit}
\end{figure}
\begin{figure}
  \resizebox{\hsize}{!}{\includegraphics{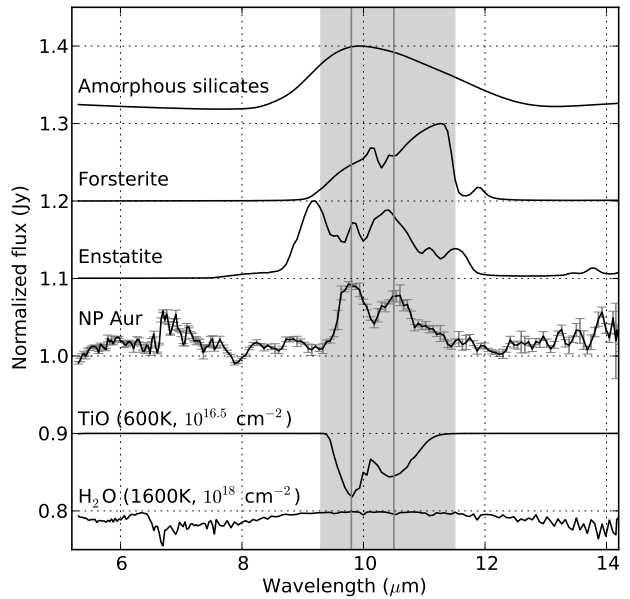}}
  \caption{Normalized Spitzer spectrum of NP~Aur, the opacities of candidate dust species and the normalized absorption of TiO and H$_2$O. The gray area and lines indicate the position of the TiO emission features.}
  \label{fig:identification}
\end{figure}

As can be seen from the full infrared Spitzer spectrum in Fig.~\ref{fig:fit}, the infrared continuum can be well reproduced by a single, 1900\,K black body. Figure~\ref{fig:identification} shows the observed spectrum, normalized to a black-body emission spectrum with a temperature of 1900\,K and an angular radius of 1.17\,mas. Above the normalized spectrum, we show the opacities of a selection of common dust species with emission features in the 10\,\um region \citep[the dust opacities were computed as a distribution of hollow spheres (DHS) as presented by ][]{Min2008}. Below the spectrum, we show the normalized absorption of TiO and H$_2$O molecules, calculated from the linelist of \citet{schwenke1998} and \citet{schwenke1997}. The region of interest in this paper is indicated by a gray band. The complex of emission lines in the 5--8\,\um range can be attributed to the emission by H$_2$O molecules, presumably in a molecular shell close above the stellar photosphere.

This figure already suggests that TiO is the only valid candidate to explain the new feature(s). The crystalline dust grains show emission features at similar wavelengths, but are always accompanied by other strong emission features that are not observed. The emission feature caused by amorphous silicates is too broad to explain the observed spectrum. Furthermore, if the 10\,\um emission were due to silicates, the spectrum should show more structure in the 18--20\,\um range.

\begin{figure}
\resizebox{\hsize}{!}{\includegraphics{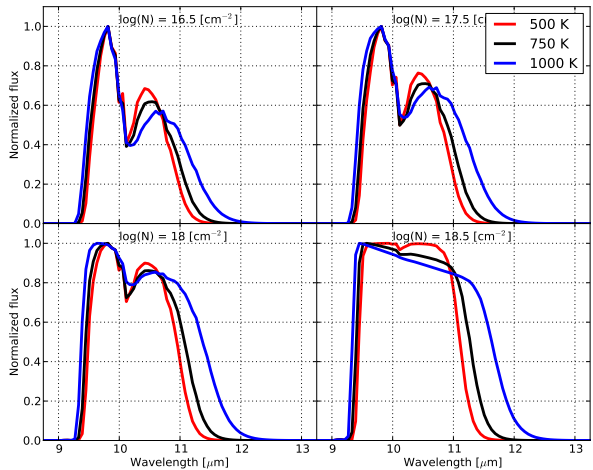}}
  \caption{Shape of the TiO emission, with the separate P and R branches visible at 9.8 and 10.3\,\um. The figures are arranged by increasing column density. The color indicates the temperature of the molecular slab.}
  \label{fig:logNandT}
\end{figure}

Figure~\ref{fig:logNandT} shows the emission feature of TiO for different column densities and different temperatures. The column densities for the models in the upper left panel correspond to optically thin emission, where the P and R branches are clearly visible as two emission features centered at 9.8 and 10.3 \,\um, respectively. For higher column densities, this difference becomes less pronounced as we move toward the optically thick regime. The lower right panel shows the optically thick TiO emission, where the P and R branches have blended into a single broad emission feature. It is clear from this figure that an increase in temperature results in a broader emission band for all column densities, while the peak of the R branch shifts from 10.3 to 10.6\,\um. An increase in temperature enhances the relative difference in peak flux between both branches.

\section{Modeling the TiO emission}
To reproduce the infrared spectrum of NP~Aur, we adopted a plane-parallel geometry and constructed a model by superimposing two extended molecular slabs above a stellar disk that provided blackbody continuum radiation. The first molecular slab contains H$_2$O, the second contains TiO. Both slabs are assumed to be isothermal, in LTE, and can be geometrically extended with regard to the stellar disk, as expected for extra-photospheric molecular layers.

As was shown in Fig.~\ref{fig:logNandT}, it is possible to constrain the temperature of the TiO molecules based on the characteristics of the emission features. However, because the column densities of TiO are close to the optically thin regime, increasing the density or the radius has approximately the same effect on the emission feature. Therefore, we cannot constrain the radius and the column density based on this TiO emission band alone. However, we can derive an upper limit on the column density of $10^{17.25}$\,cm$^{-2}$. For this upper limit, the TiO emission is optically thick. Increasing the column density above this value changes the width and shape of the emission feature significantly. This upper limit on the column density corresponds to a lower limit on the radius of the molecular layer of 4.6 R$_*$. For these limit values, the molecular layer is situated in the so-called \emph{inner wind} or extended atmosphere \citep{Cherchneff2006}.

Figure~\ref{fig:fit} shows the infrared spectrum of NP~Aur in black and the model spectrum in red. The stellar disk has a temperature of 1900\,K and a radius of 1.17\,mas. This temperature is noticeably lower than the stellar temperature, but could be explained by an optically thick layer of water that contributes to the continuum emission of the stellar disk \citep{Tsuji2000}. The H$_2$O slab has a temperature of 1600\,K, a radius of 1.5\,R$_\star$ and a column density of $10^{19.25}$\,cm$^{-2}$. For the TiO slab, we used a temperature of 600\,K and the limit values for the radius and the column density of 4.6\,R$_\star$ and $10^{17.25}$\,cm$^{-2}$, respectively. From this figure we can see that the model assuming isothermal, molecular slabs can already explain the observed emission features quite well. Adding more slabs to the model and/or adding SiO molecules did not significantly improve the goodness-of-fit.

\section{Discussion}
\label{sec:conclusions}
\subsection{The search for TiO and TiO$_2$ emission lines}
The absorption bands of TiO are one of the most prominent features in the optical spectra of cool oxygen-rich stars, for which they are often used as a temperature indicator \citep{Alvarez1997}. The rotational spectrum of TiO molecules has also been extensively studied in the laboratory at millimeter and radio wavelengths \citep{Steimle1990, Namiki1998}. But until now, the search for infrared, millimeter or radio emission of TiO molecules in the AGB wind or in molecular clouds has not yielded significant detections \citep{Churchwell1980, Millar1987, Brunken2008}. This lack of observed TiO emission may indicate that in most stars, TiO has been depleted into molecular or solid-state TiO$_2$ as soon as it was lifted above the stellar photosphere.

Assuming chemical equilibrium, \citet{Gail1998} showed that TiO$_2$ gas forms efficiently from a reaction of H$_2$O and TiO once the temperature drops below 1200\,K. Furthermore, they showed that for temperatures below 1000\,K, this reaction converts almost all TiO into TiO$_2$ gas. This TiO$_2$ gas can then be used to form solid titanium oxides, which are considered to be the primary seed grains onto which different oxides and silicates can condense \citep{Onaka1989, Tielens1990, Jeong1999}.

From our analysis, we conclude that the circumstellar environment of NP~Aur contains both H$_2$O and TiO. Under these conditions all TiO should be converted to TiO$_2$ or to solid titanium oxides \citep{Gail1998}. However, the presence of molecular TiO in the circumstellar environment of this star indicates that these reactions did not or do only partially take place. The reason for this might be the high temperatures in the shocked gas near the star. Furthermore, we compared the Spitzer spectrum to the theoretical emission spectra of TiO$_2$ grains and other titanium oxides \citep[optical constants and opacities taken from][]{Posch2003}. Although all synthetic spectra show clear emission features in the 14--20\,\um range, we did not detect any significant emission because of these dust species.

\subsection{TiO emission from oxygen-rich Mira-type S stars}
\begin{figure}
\resizebox{\hsize}{!}{\includegraphics{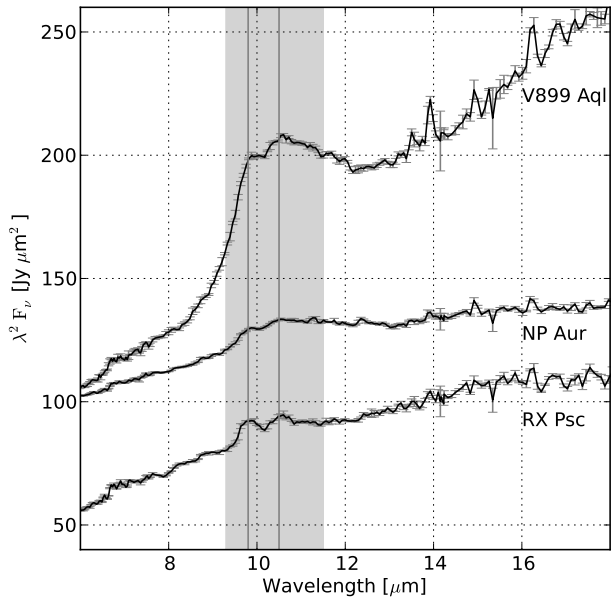}}
  \caption{Three stars in our sample with possible TiO emission features, shown in Rayleigh-Jeans units. From top to bottom: \object{V899~Aql}, \object{NP~Aur}, \object{RX~Psc}. The gray area and lines indicate the position of the TiO emission features.}
  \label{fig:candidates}
\end{figure}

Out of a sample of 87 S-type AGB stars, 49 sources do not show significant dust emission features, and of these NP~Aur is the only star that shows a clear detection of TiO. A search of the entire sample indicates that two other stars, \object{V899~Aql} and \object{RX~Psc}, show similar double peaked emission, shown in Figure~\ref{fig:candidates}. This can be explained by a superposition of TiO molecular emission on top of dust emission from silicates and alumina. 

The HERMES high-resolution spectra of NP~Aur show the Tc absorption at 4238, 4262, 4296, and 5924\,\AA. Because the half-life time of Tc is approximately $2\times10^5$ years, we can conclude that this Tc has recently been produced during a third dredge-up event. Furthermore, by comparing optical photometric and spectroscopic data to the MARCS model atmospheres calculated specifically for S-type AGB stars, it is possible to derive the stellar parameters \citep{VanEck2011}. For NP~Aur we find an effective temperature of $3200\pm300$\,K and a C/O ratio of $0.5\pm0.1$. Although the C/O ratio is one of the lowest for the stars in our sample, this is not an unusual value for MS- and S-type stars and the C/O ratio is consistent with the classification of NP~Aur as a M5:S star by \citet{stephenson2}. Furthermore, the star shows large-amplitude variations in the V-band and is classified in the General Catalog of Variable Stars as a Mira with a 334-day period \citep{GCVS}.

A comparison of the stellar properties of \object{NP~Aur} and the other stars with a tentative detection of TiO emission shows some similarities. First of all, they show only weak (or even no) dust emission, which is necessary because the contrast of the TiO emission decreases with stronger dust emission. Furthermore, all three stars are large-amplitude Mira-type pulsators, with a period of 281 days for RX~Psc and 347 days for AU~Car \citep{Smolders2012}. This is consistent with the idea that large-scale pulsations are necessary to lift the molecular layer of TiO far enough above the stellar atmosphere. Finally, both NP~Aur and RX~Psc have been classified as MS stars and are thus oxygen-rich S stars. For V899~Aql, we found no spectral classification. 

However, these pulsations and stellar characteristics resemble those of many other stars (for example AU~Car, shown in Fig~\ref{fig:candidates}) in our sample and we conclude that apart from the TiO emission features, these targets, and NP~Aur in particular, are typical oxygen-rich S-type AGB stars. Although large-amplitude pulsations, s-process enrichment and/or a low C/O ratio could be strict requirements to observe TiO in emission, they are not sufficient. A possible explanation for this might be that TiO is only visible during a short timespan, for example at a certain phase of a pulsation cycle or a short evolutionary phase.

\section{Conclusion}
We presented the detection of a new, unusual emission feature in the 9-11\,\um region. We identified this feature as the fundamental vibrational band of gaseous TiO. We showed that it is possible to reproduce this emission band using a simple isothermal, single molecular slab. This model can constrain the temperature of the TiO molecules, but because the emission comes from an approximately optically thin region, we cannot put strict constraints on the column density or outer radius of the emitting region. We discussed that this is the first time that TiO molecules are observed in the circumstellar environment of an AGB star. Based on the stellar properties of NP~Aur, the two stars with a tentative detection of this TiO emission band, and the other S stars in our sample, we argue that (i) large-amplitude pulsations, (ii) a low C/O ratio, and (iii) weak dust emission in the 9-11\,\um region might be necessary, but not sufficient requirements for the infrared spectrum to show TiO emission.

\bibliographystyle{aa}
\bibliography{references.bib}

\begin{acknowledgements}
K. Smolders, J. Blommaert, L. Decin and H. Van Winckel acknowledge support from the Fund for Scientific Research of Flanders under the grant G.0470.07. S. Van Eck is an F.N.R.S Research Associate. Based on observations made with the Mercator Telescope, operated on the island of La Palma by the Flemish Community, at the Spanish Observatorio del Roque de los Muchachos of the Instituto de Astrof\'isica de Canarias. Based on observations obtained with the HERMES spectrograph, which is supported by the Fund for Scientific Research of Flanders (FWO), the Research Council of K.U.Leuven, the Fonds de la Recherche Scientifique (FNRS), the Royal Observatory of Belgium, the Observatoire de Gen\`eve and the Th\"uringer Landessternwarte Tautenburg. This work was partly funded by an Action de recherche concert\'{e} (ARC) from the Direction g\'{e}n\'{e}rale de l'Enseignement non obligatoire et de la Recherche scientifique - Direction de la recherche scientifique - Communaut\'{e} francaise de Belgique.
\end{acknowledgements}

\end{document}